\documentclass[prd,superscriptaddress,nofootinbib,showpacs]{revtex4-1}

\usepackage{graphicx,amsmath,amsfonts,amssymb,dcolumn,epsfig,bm}

\newcommand{\bes}{\begin{subequations}}
\newcommand{\ees}{\end{subequations}}
\newcommand{\be}{\begin{equation}}
\newcommand{\ee}{\end{equation}}
\begin{document}
\title{Noise Kernel for Self-similar Tolman Bondi Metric: Fluctuations on Cauchy Horizon}
\author{Seema Satin}\email{satin@imsc.res.in } 
\affiliation{Institute of Mathematical Sciences, Taramani, Chennai, India }
\author{Kinjalk Lochan} \email{ kinjalk@tifr.res.in}
\affiliation{Tata Institute of Fundamental Research, Colaba, Mumbai, India }
\author{Sukratu Barve} \email{sukratu@cms.unipune.ac.in }
\affiliation{Center for Modeling and Simulation, University of Pune, Pune, India}


\begin{abstract}
We attempt to calculate the point separated Noise Kernel for self similar Tolman Bondi metric, using a method similar to that developed by  
Eftekharzadeh et. al for ultra-static spacetimes referring to the work by Page. In case
of formation of a naked singularity, the Noise Kernel thus obtained is found to be regular except on the Cauchy 
horizon, where it diverges. The behavior of the noise in case of the formation of a covered singularity is found to be regular. 
This result seemingly renders back reaction non-negligible which questions the stability of the results obtained from the semiclassical treatment 
of the self similar Tolman Bondi metric.
  
\end{abstract}


\maketitle

\section{Introduction}

Cosmic censorship ever since its proposition in 1969 has been an issue studied
mostly within the framework of classical relativity. This includes important
conjectures like the hoop conjecture, studies on stability of scalar fields on
spherically symmetric collapse metrics as well as particular examples
of gravitational collapse metrics. Some studies have addressed gravitational
collapse at the semiclassical level though most of the work concerns 1+1 dimensional metrics.  In the semiclassical theory \cite{birrel,wald}, the 
quantum expectation of the energy momentum tensor (also referred to as RSET or renormalized stress energy tensor) is central to studying  
quantum fields on spacetimes. One line of investigation in applying this to
 gravitational collapse is to question the features like trapped surfaces
 which arise earlier and the final fate of the collapse \cite{visser} \cite{mattingly2}. Another line is to work out the quantum stress tensor
when one employs as a background spacetime the classical relativistic examples
of gravitational collapse and study its effects. In the later in particular, the metric representing pressureless
fluid collapse in spherical symmetry, the Lemaitre Tolman Bondi (LTB) metric is a common choice. We compare with this line of investigation in this work.

The classical Lemaitre Tolman Bondi (LTB) metric exhibits both naked and covered singularities which can be easily related to  data on the initial Cauchy surface. 
The transition from covered to naked singularities appears to occur on account 
of inhomogeneities in initial data. Further, at the semiclassical level it 
appears that this increasing inhomogeneity causes a drastic change in the 
behavior of the
 quantum stress tensor. In fact, it is found that the quantum stress tensor 
diverges in the naked singular case where the Cauchy horizon is expected to 
 form whereas there is no divergence anywhere in the covered singularity case.
This divergence is physically interpreted as an energy burst on the Cauchy 
horizon \cite{Sukratu}.

 It is still not clear if such a behavior is generic as the results obtained
 are in fact exact in 1+1 dimensional cases and for a few cases of
 background metrics. Also the calculations involve a background metric which remains 
 unaffected by the quantum stress tensor. In other words, the backreaction
 is not implicitly incorporated in the results. However, physical arguments 
are offered for the divergence which rely on high curvature regions of 
spacetime causing particle production. Such regions would not be exposed in 
the covered case as against the naked case and it is {\it likely} that the 
semiclassical divergence is actually a general feature. 

 Semiclassical gravity is however fraught with its share of interpretational
 issues apart from  operational difficulties like backreaction. Since the idea 
 by Leon Rosenfeld of using the quantum stress tensor on the right of the 
 Einstein field equations, several issues have been raised
 \cite{ mattingly}. At least some of them involve serious objections to employing
 an averaged quantity for the stress tensor. Indeed one can imagine that
 in situations where fluctuations are important, the use of the 
 semiclassical quantum averaged stress tensor would not be justified for any
 physical interpretation like the one above. So for a complete consideration it looks
 natural to address the fluctuations as well, whenever they can be argued to be significant.
 
Stochastic gravity seeks to remedy this situation, 
addressing the effect of including fluctuations of the energy momentum tensor 
and possibly its backreaction on the metric as well. This is accomplished 
using the  Einstein Langevin Equation \cite{Hu1,Verd2}. 
The central object in the calculations of stochastic gravity is the noise
Kernel which gives an idea of the stochastic source, that being in addition
to the quantum stress tensor source.
The Noise Kernel is the vacuum expectation value of symmetrized stress-energy 
bi-tensor for a quantum field in curved spacetime \cite{Verd3,Hu2}. It characterizes fluctuations
of stress tensor which play as mentioned above, a 
significant role in analysis of quantum effects in curved spacetime. These 
fluctuations lead to fluctuations in the metric, themselves understood in two parts as \textit{induced}
or \textit{passive} fluctuations \cite{Hu1}. Induced fluctuations have played 
an important role in many studies involving backreaction. Backreaction 
problems in gravity and cosmology for example have been addressed
 \cite{Sukanya} using Einstein-Langevin equations.

This far the effect of quantum fluctuations has not been studied in
 the context of gravitational collapse, the physical scenario for studying 
cosmic censorship. This is firstly because of the
operational difficulty in calculating the central quantity, the Noise Kernel 
which gives an idea of the stochastic source term. Secondly, the quantum
 stress tensor does not add significantly to the classical source for
most of the spacetime regions. This would support any interpretation
ignoring backreaction effects of the quantum stress especially regarding the 
effects on quantum fields like pair creation or Hawking radiation (at `late' times) \cite{FredenhagenHaag}   
However, it is physically important to assure that the quantum averaged
source is indeed a good approximation to the quantum field contribution to 
the source. Any effect of fluctuations, which appear as the stochastic
source described above would have to be negligible.
However, it has been a challenge in stochastic gravity to come up with a method to 
calculate the Noise Kernel for specific background metrics. There are various 
approaches that have been used for such calculations, in the case of 
Minkowski, de-Sitter, 
anti-deSitter and Schwarzschild Spacetimes  in the coincidence limit \cite{Nadal,Fleming,Hu4,Roura,Roura2,Cho}.
 Recently, Eftekharzadeh et. al. \cite{Ifte} have 
devised a method to compute Noise Kernel for conformally invariant quantum fields in Schwarzschild
spacetime under the Gaussian approximation. This is based on method 
developed by Page \cite{Page} and provides a possibly general way to estimate 
the Noise Kernel approximately.

We seek to apply stochastic gravity 
techniques in the study of cosmic censorship to the gravitational collapse scenario, particularly to the LTB metric. This metric
 has been hitherto studied only at the level of quantum fields on spacetime. It has been demonstrated that stress energy tensor 
diverges at the Cauchy horizon, despite Cauchy horizon being perfectly regular \cite{Sukratu}. Such results do not take into
account backreaction of the quantum fields on spacetime. The motivation of calculating 
Noise Kernel in LTB background is to test the validity of the above mentioned result in view of backreaction.
Our aim in carrying out this work has been to investigate this in the context of gravitational collapse and suggest 
interpretations of the particular calculations for the Noise Kernel.
In this paper, we employ a similar method to that of \cite{Ifte} and compute the
Noise Kernel on self similar Tolman Bondi metric background for conformally 
invariant fields under Gaussian approximation.   
We obtain the Noise Kernel in this case which diverges at the Cauchy horizon, seemingly making backreaction important for consideration and hence
questioning any attempt to study the spacetime in a semi-classical or perturbative manner. 

The paper is organized as follows.
In section II we introduce the building blocks of stochastic gravity, the 
Noise Kernel and the required Wightman function. 
Section III describes the conformal rescaling of the Tolman Bondi metric and its conformally equivalent ultra-static form is obtained.
We obtain the expansions of Synge function and Van Vleck Morette determinant in the coordinates of this ultra-static metric. In section IV
we relate the Noise Kernel of this metric to that of self-similar Tolman Bondi. We argue here for a plausible divergence.
In the end we summarize the  results and have a discussion of interpretation of the divergence.

\section{Noise Kernel in ultra-static spacetimes}

The Noise Kernel is introduced briefly and related to the contribution
of stress fluctuations to the source in Einstein equations. The nature
of the assumed quantum state is described for this situation. Its relation
to the Noise Kernel is mentioned and its expression is provided in case of 
an ultra-static spacetime.

\subsection{ Einstein Langevin Equation and Noise Kernel}
The Noise Kernel for quantized matter field is given by
\be
N_{abc'd'} (x,x') = \frac{1}{2} \langle \{ \hat{t}_{ab}(x), \hat{t}_{c'd'}(x')
 \} \rangle 
\ee
where
\[ \hat{t}_{ab}(x) \equiv \hat{T}_{ab}(x) - \langle \hat{T}_{ab}(x) \rangle \]
and $< \dots >$ is the quantum expectation value taken with respect to a normalized
state. $\hat{T}_{ab}$ is the stress tensor operator of the field.
The classical stress tensor of a conformally invariant scalar field $\phi$ is given by
\be 
T_{ab} = \nabla_a \phi \nabla_b \phi - \frac{1}{2} g_{ab} \nabla^c \phi
\nabla_c \phi + \frac{1}{6}( g_{ab} \Box - \nabla_a \nabla_b + G_{ab} ) \phi^2
\ee
As fit for doing quantum field theory on curved space-time, the field $\phi$  is raised to
the level of an   operator whereas $g_{\mu \nu}$ is to be treated classically.

The Noise Kernel embodies the contributions of the higher correlation functions
in the quantum field on account of which it may be used to interpret issues
related to the quantum nature of space-time. Two point functions of energy 
momentum tensor would involve fourth order correlations of the quantum field,
for example, which would affect the coherence of the geometry\cite{kineticqg}, if we were to 
employ the Einstein Langevin equations (often referred to as semiclassical
Einstein Langevin equations)
\be
G_{\mu \nu} = < \hat{T}_{\mu \nu}> + \hat{\xi}_{\mu \nu},
\ee
where $\hat{\xi}_{\mu \nu}$ is a random variable (tensor) which is closely
related to the Noise Kernel. The relations 
\begin{eqnarray}
 <\hat{\xi}_{\mu \nu} (x) >_s &=& 0 \nonumber\\
 < \hat{\xi}_{\alpha \beta}(x) \hat{\xi}_{\mu'\nu'}
(x')>_s &=& 
N_{\alpha \beta \mu' \nu'}(x,x') 
\end{eqnarray}
completely characterize the variable $ \hat{\xi}_{\alpha \beta}(x)$.
The statistical expectation $< \hspace{0.25 in}>_s $ is taken over various stochastic  realizations of
 the Gaussian source $\hat{\xi}_{\mu \nu} (x) $. Each of the realizations 
would lead to a metric solution
\be \label{eq:gret}
h_{\alpha \beta} (x) = h^{(0)}_{\alpha \beta} (x) + 8 \pi \int d^4 y' 
\sqrt{-g(y')} G^{(ret)}_{\alpha \beta \gamma' \delta'} (x,y') \xi^{\gamma' \delta'} (y')
\ee
where $h^{(0)}_{\alpha \beta}$ is the solution to the semiclassical equation
\[
G_{\mu \nu} = < \hat{T}_{\mu \nu}>
\]
and $ G^{(ret)}_{\alpha \beta \gamma' \delta'}$ serves as the retarded 
propagator of the semiclassical Einstein Langevin equations with vanishing
initial conditions.
The stochastic realization would contain information about coherence of 
geometry, $ \hat{\xi}^{\mu \nu} (x)$ being treated as a random variable.  

\subsection{The quantum state for the Noise Kernel: Wightman function}

Before interpreting physically any particular noise calculation, the nature of 
the quantum state of the field needs to be taken into account. The state we 
consider in our calculations turns out to be Hadamard in addition to
our assumptions of it being quasi-free and thermal. We discuss these three
qualifiers of the state below. 

Firstly, the Hadamard nature of the state guarantees that the stress tensor is
well defined within the maximal Cauchy development and obeys the Wald axioms \cite{wald}. 
We will then analytically continue the state across the Cauchy horizon. This consideration can be straightforwardly generalized 
to Schwarzschild spacetime as considered in \cite{Ifte}.
Operationally, we are not interested in the stress tensor and would 
not require any coincidence limits, even for the Noise Kernel for addressing issues of fluctuations.
However, the fluctuations obtained would be specific
to the quantum state and a physically reasonable choice of state is warranted.
Hence the Hadamard nature of the state is accorded physical significance in
the discussion of our results.

In our calculations below, we have emphasized the singularity structure of the 
symmetrized form of the two point function we use. That being represented by a 
Hadamard expansion, we conclude that the quantum state being used is Hadamard. 
The expressions we use are for a general quasi-static spacetime and based
on the approximation carried out by Page and so the conclusion would be
more general. 
However, a careful interpretation of the divergent expression in $1/\sigma$ with 
$\sigma^{2}$ being the Synge function, is required to support the claim that 
the state is Hadamard. Such an interpretation in the sense of distributions is 
provided by Kay and Wald \cite{kaywald}. A rigorous definition is provided by Radzikowski \cite{radzikowski}
based on micro-local analysis but we do not attempt to relate to that in this paper. 
Secondly, we operationally assume that the Wightman two point function 
defined as
\[ {\cal G}(x,x') = \langle \phi(x) \phi(x') \rangle     \]
determines the quantum state of the field. Indeed, that is true if one is 
working with quasi-free (or Gaussian) states. Quasi-free states have been 
extensively used \cite{wald} in curved spacetime. They conveniently lead to 
Fock space representations of one particle Hilbert spaces via the GNS 
construction on the Weyl algebra of observables. One advantage of this is 
that can admit a concept of particles in a limited manner, despite having 
curved spacetime at hand. For example, large fluctuations implied by the
Noise Kernel could be interpreted as highly fluctuating particle creation.

Thirdly, it is assumed that the state is thermal or K.M.S. with temperature
$\kappa$.  It turns out that the Wightman function can be approximated
for thermal states in a particular manner called Gaussian approximation.
We employ the related expression in the last subsection for ultra-static 
spacetimes.

Our results would thus be applicable to a quasi-free thermal Hadamard state
of a conformally invariant field. The state specific information is not 
contained in the singularity structure. So changing to a different 
quasi-free Hadamard state would amount to adding an analytic function to 
the symmetrized two point function. We find that the Noise Kernel expression 
would amount to adding an analytic function to the symmetrized
two point function. We find that the Noise Kernel expression would
not change at the $\kappa^{0}$ level. The divergence on Cauchy horizon,
being at all orders in $\kappa$, is expected  to persist for any
quasi-free Hadamard state.

\subsection{Noise Kernel in terms of the Wightman function}

The properties and the expressions of a Noise Kernel, given a
quasi-free (or Gaussian) state of quantum matter field, are obtained in
\cite{hu11}. Further, for conformally invariant fields the Noise Kernel will 
be given are given in terms of the Wightman function as \cite{Ifte}

\begin{equation}
 N_{abc'd'}  =
 {\rm Re} \left\{  \bar K_{abc'd'}
  + g_{ab}   \bar K_{c'd'}
 + g_{c'd'} \bar K'_{ab}
 + g_{ab}g_{c'd'} \bar K \right\}
\label{general-noise-Kernel}
\end{equation}

where
\bes
\label{General-Noise}
\begin{eqnarray}
9  \bar K_{abc'd'} &=&
%
4\,\left( {\cal {\cal G{}}}\!\,_{;}{}_{c'}{}_{b}\,{\cal G{}}\!\,_{;}{}_{d'}{}_{a} +
    {\cal G{}}\!\,_{;}{}_{c'}{}_{a}\,{\cal G{}}\!\,_{;}{}_{d'}{}_{b} \right)
%
+ {\cal G{}}\!\,_{;}{}_{c'}{}_{d'}\,{\cal G{}}\!\,_{;}{}_{a}{}_{b} +
  {\cal G}\,{\cal G{}}\!\,_{;}{}_{a}{}_{b}{}_{c'}{}_{d'} \cr
%
&& -2\,\left( {\cal G{}}\!\,_{;}{}_{b}\,{\cal G{}}\!\,_{;}{}_{c'}{}_{a}{}_{d'} +
    {\cal G{}}\!\,_{;}{}_{a}\,{\cal G{}}\!\,_{;}{}_{c'}{}_{b}{}_{d'} +
    {\cal G{}}\!\,_{;}{}_{d'}\,{\cal G{}}\!\,_{;}{}_{a}{}_{b}{}_{c'} +
    {\cal G{}}\!\,_{;}{}_{c'}\,{\cal G{}}\!\,_{;}{}_{a}{}_{b}{}_{d'} \right)  \cr
%
&& + 2\,\left(
{\cal G{}}\!\,_{;}{}_{a}\,{\cal G{}}\!\,_{;}{}_{b}\,{R{}_{c'}{}_{d'}} +
    {\cal G{}}\!\,_{;}{}_{c'}\,{\cal G{}}\!\,_{;}{}_{d'}\,{R{}_{a}{}_{b}} \right)  \cr
%
&& - \left( {\cal G{}}\!\,_{;}{}_{a}{}_{b}\,{R{}_{c'}{}_{d'}} +
  {\cal G{}}\!\,_{;}{}_{c'}{}_{d'}\,{R{}_{a}{}_{b}}\right) {\cal G}
%
 +{\frac{1}{2}}  {R{}_{c'}{}_{d'}}\,{R{}_{a}{}_{b}} {{\cal G}^2}
\end{eqnarray}
\begin{eqnarray}
 36  \bar K'_{ab} &=&
%
8 \left(
 -  {\cal G{}}\!\,_{;}{}_{p'}{}_{b}\,{\cal G{}}\!\,_{;}{}^{p'}{}_{a}
 + {\cal G{}}\!\,_{;}{}_{b}\,{\cal G{}}\!\,_{;}{}_{p'}{}_{a}{}^{p'} +
  {\cal G{}}\!\,_{;}{}_{a}\,{\cal G{}}\!\,_{;}{}_{p'}{}_{b}{}^{p'}
\right)\cr &&
%
+ 4 \left(
    {\cal G{}}\!\,_{;}{}^{p'}\,{\cal G{}}\!\,_{;}{}_{a}{}_{b}{}_{p'}
  - {\cal G{}}\!\,_{;}{}_{p'}{}^{p'}\,{\cal G{}}\!\,_{;}{}_{a}{}_{b} -
 {\cal G}\,{\cal G{}}\!\,_{;}{}_{a}{}_{b}{}_{p'}{}^{p'}
\right) \cr
%
&& - 2\,{R'}\,\left( 2\,{\cal G{}}\!\,_{;}{}_{a}\,{\cal G{}}\!\,_{;}{}_{b} -
    {\cal G}\,{\cal G{}}\!\,_{;}{}_{a}{}_{b} \right)  \cr
%
&&  -2\,\left( {\cal G{}}\!\,_{;}{}_{p'}\,{\cal G{}}\!\,_{;}{}^{p'} -
    2\,{\cal G}\,{\cal G{}}\!\,_{;}{}_{p'}{}^{p'} \right) \,{R{}_{a}{}_{b}}
%
 - {R'}\,{R{}_{a}{}_{b}} {{\cal G}^2}
\end{eqnarray}
\begin{eqnarray}
 36 \bar K &=&
2\,{\cal G{}}\!\,_{;}{}_{p'}{}_{q}\,{\cal G{}}\!\,_{;}{}^{p'}{}^{q}
+ 4\,\left( {\cal G{}}\!\,_{;}{}_{p'}{}^{p'}\,{\cal G{}}\!\,_{;}{}_{q}{}^{q} +
    {\cal G}\,{\cal G{}}\!\,_{;}{}_{p}{}^{p}{}_{q'}{}^{q'} \right)  \cr
&& - 4\,\left( {\cal G{}}\!\,_{;}{}_{p}\,{\cal G{}}\!\,_{;}{}_{q'}{}^{p}{}^{q'} +
    {\cal G{}}\!\,_{;}{}^{p'}\,{\cal G{}}\!\,_{;}{}_{q}{}^{q}{}_{p'} \right)  \cr
&& + R\,{\cal G{}}\!\,_{;}{}_{p'}\,{\cal G{}}\!\,_{;}{}^{p'} +
  {R'}\,{\cal G{}}\!\,_{;}{}_{p}\,{\cal G{}}\!\,^{;}{}^{p} \cr
&& - 2\,\left( R\,{\cal G{}}\!\,_{;}{}_{p'}{}^{p'} +
{R'}\,{\cal G{}}\!\,_{;}{}_{p}{}^{p} \right)
     {\cal G}
+  {\frac{1}{2}} R\,{R'} {{\cal G}^2}  \;
\end{eqnarray}
\label{generalnoise}
\ees
are given in terms of the Wightman function. Primes on indices represent the point $x'$ while the entities at $x$ are the unprimed ones. 
$R_{ab}$ is the Ricci tensor and $R$ is the Riemann curvature.
\newline
\subsection{Wightman function in ultra-static spacetime}
Next, we give a short review of the method used in \cite{Ifte} for ultra-static
metric. The metric in a static spacetime takes the form 
\be
ds^2 = g_{\tau \tau}(\vec{x}) d \tau^2 + g_{ij}(\vec{x}) dx^idx^j
\ee
This can be  transformed into an ultra-static form, called the optical metric,
by a conformal transformation. This optical metric takes the following form
\be ds^2 = dt^2 + g_{ij}(\vec{x}) dx^i dx^j \ee
where the metric functions $g_{ij}$ are independent of time $t$. The Synge function
(half of the square of the proper distance of the shortest geodesic between two points) can thus in ultra-static space-time 
take the form
\[\sigma(x,x') = \frac{1}{2} ((t-t')^2 -\mathbf{r })^2 \] 
where $\mathbf{r }^{2}$ is twice the spatial part of the Synge function and it depends
only on spatial co-ordinates as in \cite{Ifte}. 

 The Wightman function in an optical background metric for a thermal (K.M.S.) 
state  can be calculated under the Gaussian approximation \cite{Ifte}. This 
approximation was first carried out by Page using the Schwinger De-Witt 
expansion in the context of calculating quantum stress tensors. In this scheme
\be
{\cal G}(\delta t, \vec{x}, \vec{x}') = \frac{ \kappa \sinh \kappa \mathbf{r}}{
8 \pi^2 \mathbf{r}[\cosh(\kappa \mathbf{r}) - \cosh(\kappa \delta t)]}
U(\delta t, \vec{x},\vec{x'}), \label{Wightman}
\ee
where 
\[ \mathbf{T} = \frac{\kappa}{2 \pi} \]
is the temperature of the  thermal state considered\footnote{In Schwarzschild spacetime, the $\kappa$ has been chosen to
be the surface gravity on the event horizon null surface. However, one can work with  a general thermal state without
ascribing this significance to it.} and\\
\[ \delta t= t-t'\]
while
\begin{eqnarray}
U(x,x') & \equiv & \Delta^{1/2} (x,x')  \label{Ufunc} \\
\Delta(x,x') & \equiv &  \frac{1}{\sqrt{-g(x)}\sqrt{-g(x')}} det(\sigma ;_{
ab'}) \label{Del}
\end{eqnarray}
where $\Delta(x,x')$ is known as the Van Vleck Morette determinant.

\section{Conformally Optical form of Self Similar Tolman Bondi metric}
\noindent In order to evaluate the Noise Kernel for Schawrzschild spacetime  Eftekharzadeh et.al. \cite{Ifte} have used an approach of conformally
rescaling the metric to the form of an ultra-static 
 (optical) metric. First the Noise Kernel is evaluated in this ultra-static spacetime. Then the Noise Kernel of 
 Schwarzschild spacetime is obtained by just rescaling back the result in the optical case. We will be applying the similar 
 technique to self-similar Tolman Bondi(TB) metric, which is given by the line element
\be
ds^2 = dt^2 - R' dr^2 - R^2 d \theta^2 - R^2 \sin^2 \theta d \phi^2
\ee
where $R(t,r)$ is the area radius
\be
R^{3/2}(t,r)  =  r^{3/2}(1 - \frac{3}{2} \frac{t}{r} \sqrt{\lambda})
\ee
is characterized by a dimensionless parameter $\lambda$. 
The collapse rate can also be given as in terms of this parameter
\be
\dot{R} = - r \sqrt{\frac{\lambda }{R}}.
\ee
The initial data for this metric is regular and a curvature singularity gets eventually formed
at $r=0$. The singularity is naked if $ \lambda < 1/8 $ and is covered otherwise. 

For conformal transformation is it useful to work in a new coordinate system $(t,z,\theta,\phi)$ where
$ z = t/r $. Then
\begin{eqnarray}
R^{3/2}(z) & = & r^{3/2}(1 - \frac{3}{2} z \sqrt{\lambda}) \\
R'(z)& = & \frac{ 1 - \frac{z}{2} \sqrt{\lambda}}{(1-
 \frac{3 z \sqrt{\lambda}}{2})^{1/3}} \\
\end{eqnarray}
The self similar TB metric can now be put in the conformally ultrastaic form as
\be \label{eq:metric} 
ds^2 = \Omega^2 [dT^2 - f_1(z)^2  dz^2 -  z^2 f_2(z)^2 d \Omega^2 ],
\ee
 where
 \begin{eqnarray}
  \Omega^2=\left(1 - \frac{R'^2}{z^2} \right) t^2\\
  f_1(z)^2 = \left(\frac{ R'}{z^2 - R'^2}\right)^2\\
  f_2(z)^2 = \frac{ (1-\frac{3}{2} z \sqrt{\lambda})^{4/3}}{z^2(z^2 - R'^2)}, 
 \end{eqnarray}
 and \\
\[dT=\frac{dt}{t}+\frac{R'^{2}}{z(z^{2}-R'^{2})}dz.\]
Since the metric (\ref{eq:metric}) is in conformally related to an ultra-static metric, the Synge function can readily be obtained in the
following form as is done in \cite{Ifte} for Schwarzschild metric in optical 
form. We apply the same procedure for TB metric in the optical form.
The Synge function expansion used here becomes 
\[ \sigma = \sum_{ijk} s_{ijk}(z) \delta T^{2i} \eta^j \delta z^k \]
where $$ \delta T = (T-T'),$$
$$ \delta z = (z-z'),$$
$$ \eta+1 = \cos(\theta)\cos(\theta') +\sin(\theta) \sin(\theta') \cos(\phi - \phi').$$
The expression for Synge function is obtained as
\begin{eqnarray}
\sigma (x,x') & = & \frac{1}{2}[ \delta T^2 - \delta z^2 f_1(z)^2+ 2\eta z^2 f_2(z)^2\\ \nonumber
& & -2 \delta z \eta ( z f_2(z)^2 + z^2 f_2(z) f_2'(z)) +\\  \nonumber
& &   \delta z^3 f_1(z) f_1'(z)]+O[(x-x)^4]. 
\end{eqnarray}
The function $U(x,x')$ present in the Wightmann function (\ref{Wightman}) can be expanded in powers of $(x-x')$ using the Synge function
(\ref{Ufunc}),(\ref{Del}) as follows:
\begin{eqnarray}
U(x,x')& = &  1+ \delta z^2 \left[\frac{f_1'(z)}{6 z f_1(z)}-
\frac{f_2'(z)}{3 z f_2(z)}+
\frac{f_1'(z) f_2'(z)}{6 f_1(z) f_2(z)}\right] \nonumber \\  
& & - \delta z^2\left[ \frac{f_2''(z)}{6 f_2(z)}\right]
 -\eta\left[\frac{1}{6}-\frac{f_2(z)^2}{6 f_1(z)^2}+
 \frac{z f_2(z)^2 f_1'(z)}{6 f_1(z)^3}-
 \frac{2 z f_2(z) f_2'(z)}{3 f_1(z)^2}\right]  \nonumber \\ 
 & & -\eta\left[\frac{z^2 f_2(z) f_1'(z) f_2'(z)}{6 
 f_1(z)^3}-\frac{z^2 f_2'(z)^2}{6 f_1(z)^2}-
 \frac{z^2 f_2(z) f_2''(z)}{6 f_1(z)^2}\right] \nonumber\\
 & & + O[(x-x')^3]
 \end{eqnarray}
 With $U$ and the Synge function obtained for the optical self-similar TB spacetime we can now evaluate the 
 Wightman function which will be used to obtain the Noise Kernel.

\section{ Noise Kernel expression for (self similar) Tolman Bondi  spacetime}
The  Wightmann function for a thermal state with temperature $\kappa$ in the
 Gaussian approximation on ultra-static spacetime is given by \cite{Ifte}
\begin{eqnarray}
{\cal G}(x,x') & = & \frac{1}{8 \pi} [ \frac{1}{\sigma} + \frac{\kappa^2}{6} - 
\frac{\kappa^4}{180} (2 \delta T^2 + \sigma) + O[(x-x')^4]] U(x,x')
\end{eqnarray}
 
\footnote{At this point we note that the $U$ above is analytic in its domain and
the prefactor contains the singularity structure. The same singularity 
structure would present in the symmetrized two point function. 
 We have a Hadamard singularity structure above, in fact without
 any $\log \sigma$ term.}
 The Noise Kernel, when points are separated can now be computed using  the 
above Green's function, in usual way as given in  \cite{Ifte}. Here we will present one component of the
Noise Kernel for demonstrating some important results for our metric. 
The  expression  for Noise Kernel(component given by $N_{TTT'T'}$ for points
separated in
$T$, where $ \delta z = \eta = 0$ and $ \delta T \neq 0 $ .
The noise Kernel $ N_{TTT'T'}  $ can be expanded in terms of various powers of $\kappa$.
These coefficients in such an expansion are presented in the tabular form below : 
\newpage
\begin{table}
\begin{tabular}{|c|c|}
\hline
                                                   &       \\   
$\text{Coefficient of}\hspace{0.25 cm}  \kappa^0 $ & $ \frac{ z^4 f_1^2}{R'^2 t^2 (t - \delta t)^2} 
                                  \text{\large\{} \frac{13}{12 \delta T^8 \pi^4}+[-4 f_1^4+3 z^2 f_2^2 f_1'^2+2 zf_1 f_2\{-2 z f_1' f_2'+
                                  f_2$ \\ \\
                                &  $ (-2 f_1'+z f_1'')\}+4 f_1^2 \{2 f_2^2+ 2 z^2 f_2'^2+z f_2 (6 f_2'+z f_2'')\}]/[72 \delta T^6 \pi ^4 z^2 f_1^4 
                                 f_2^2] $\\ \\ 
                                & $+[27 f_1^7+50 z f_1^4 f_2 f_1' (f_2+z f_2')+  360 z^3 f_2^3 f_1'^3 (f_2+ zf_2')$ \\  \\
                                
                                & $-10 f_1^5 [3 f_2^2+ 3 z^2 f_2'^2 +z f_2 (16 f_2'+5 z f_2'')]-5 z^2 f_1 f_2^2f_1' (z^2 f_1' f_2'^2 + $\\ \\
                                & $ f_2^2 f_1'+ 48 z f_1'')+  2 z f_2 [24 z f_2'f_1''+f_1' (73 f_2'+36 z f_2'')]]+$\\ \\
                                & $ 2 z f_1^2 f_2 \{-49 z^3 f_1'f_2'^3+z^2 f_2 f_2' [12 z f_2'f_1''-f_1' (209 f_2'+ $\\ \\
                                & $ 31 z f_2'')]+f_2^3 [-49 f_1'+12 z (f_1''+z f_1^{(3)})] + z f_2^2 [12 z \{4 z f_1'' f_2''+ $\\ \\
                                & $f_2' (10 f_1''+z f_1^{(3)})\}+ f_1' [-209 f_2'+z (185 f_2''+72 z f_2^{(3)})]]\}+f_1^3 \{3 f_2^4+$\\ \\
                                & $3 z^4 f_2'^4+2 z^3 f_2f_2'^2 (104 f_2'+  49 z f_2'')+ z^2 f_2^2 [678 f_2'^2+67 z^2 f_2''^2-8 z f_2'(-49 f_2''+$ \\ \\
                                & $3 z f_2^{(3)})]+ 2 z f_2^3 [104 f_2'+z (13 f_2''-12 z (5 f_2^{(3)}+ z f_2^{(4)})]\}/[25920 \delta T^4 \pi ^4 z^4 f_1^7 f_2^4]\text{\large\}} $\\ \\
\hline
                                                   &       \\   
$\text{coefficient of}\hspace{0.25 cm}  \kappa^2 $ &  $\frac{ z^4 f_1^2}{R'^2 t^2 (t - \delta t)^2} \text{\large \{}\frac{5}{72\delta T^6 \pi ^4} +[2 f_1^4+9 z^2 
                                                       f_2^2 f_1'^2+ 2 z f_1 f_2 \{8 z f_1' f_2'+ f_2 (8 f_1'+3 z f_1'')\}$\\ \\
                                                   &$+ 2 f_1^2 \{5 f_2^2+ 5 z^2 f_2'^2-2 z f_2 (3 f_2'+4 zf_2'')\}]/[2592 \delta T^4 
                                                   \pi ^4 z^2 f_1^4 f_2^2] \text{\large \}}$\\ \\
\hline
                                                   &       \\   
$ \text{coefficient of}\hspace{0.25 cm}  \kappa^4 $  & $\frac{z^4 f_1^2}{ R' t^2(t- \delta t)^2}[-\frac{53}{1080 
\delta T^4 \pi ^4}] + O[(x-x')^3]$\\ \\
\hline
\end{tabular}
\end{table}
\newpage
The above expansion for the Noise Kernel has been obtained after conformal transformation of the optical form back to the original 
form (\ref{eq:metric}). 
We have  displayed the expression of $N_{TTT'T'}$ for points separated in $T$ co-ordinate only. This
  turns out to be interesting as it yields a divergence at the Cauchy
  horizon with leading orders in separation. This takes place despite the 
Cauchy horizon itself being regular in that there is no curvature singularity 
along it. 
\subsection{Structure for naked singularity and comparison with covered case}
 In case of the naked singular solution,
 the Cauchy Horizon is given by the smaller root $z_{-}$ of $(z^2 - R'(z)^2)$ for 
 self similar TB metric \cite{Sukratu}. This leads to divergence of the factors
 $f_1(z)$ and $f_2(z)$ if evaluated on the Cauchy Horizon. We could choose $T$
 for the point on the Cauchy Horizon and $T'$ for the one elsewhere. The 
 behavior thus obtained would be clearly of the point separated Kernel.

 The terms for various orders of $\kappa$ displayed above
 diverge  on the account of the appearance of $f_1(z)$ in the factor (the appearance of $f_1(z)\sim\frac{1}{z-z_{-}}$ 
 and $f_2(z)\sim\frac{1}{\sqrt{z-z_{-}}}$
 occurring in terms in the brackets essentially does not contribute any singular structure) leading to our main
 result, the divergence of the Noise Kernel on the Cauchy Horizon.
It should be noted that the above divergence has been obtained using metric (\ref{eq:metric}). We have been constrained by our approach to use such a metric
in coordinates ill behaved at the Cauchy horizon
 \footnote{ The metric in the earlier ($t,r,\theta,\phi$) co-ordinate system 
is perfectly regular on Cauchy horizon and we have conformally rescaling back our results. Our results would thus hold for any other metric related to it by 
regular coordinate system transformations.}.
Suitable co-ordinate transformations can be used to remedy this. From the metric one can observe that the
coordinate transformation factors acting on $N_{TTT'T'}$ above would need to
diverge at the Cauchy Horizon as $\frac{1}{z-z_{-}}$ similar to \cite{Sukratu}.
The divergence of the expression of the Noise Kernel above is rather enhanced
if the metric is demanded to be regular at the Cauchy horizon.

 In the case of covered singularity $(z^2 - R'(z)^2)$ has no real roots.
 So the factors of $f_1(z)$ and $f_2(z)$ do not yield divergences. Nor do the
 rest of the factors. The metric used is also regular everywhere.

 \subsection{Singularity structure}
 
The issue of regularity also occurs in the noise expressions of \cite{Ifte}
as the Schwarzschild metric used is expressed in the usual coordinates
rather than say Kruskal coordinates at the event horizon. For interpreting
the expressions on the event horizon, further coordinate transformations 
would be necessary on the Noise Kernel.

The above analysis is restricted to the Noise Kernel with only $\delta T \ne 0$.
We obtain a similar divergence on Cauchy Horizon (in the naked singular case)
for $\delta z \ne 0$ separation as well. The corresponding noise expressions
however are not short enough for explicit display in this communication.

Our results above for the Noise Kernel would be applicable to a 
particular quasi-free thermal Hadamard state of a conformally invariant 
field on self similar LTB spacetime. It would be of interest to examine
if this could be applicable to any quasi-free thermal Hadamard state. 

The state specific information is not contained in the singularity structure. 
So changing to a different quasi-free Hadamard state would amount to adding an
analytic function to the symmetrized two point function. If the above expressions of the Noise Kernel are examined, we find that the Noise Kernel expression 
would not change at the $\kappa^{0}$ level. The divergence on Cauchy horizon (in the naked singular case) being at all orders in $\kappa$, is expected  to 
persist for any quasi-free Hadamard state.

\section{Results and Discussion}

We have thus obtained the Noise Kernel for a conformally coupled scalar
field in the self similar Tolman Bondi metric in an approximate form for
 a quasi-free thermal Hadamard state. We seek to interpret this result below. 

We know that the Noise Kernel affects the induced fluctuations of the 
metric. In equation (\ref{eq:gret}), however the $G^{ret}$ function required
would need a background metric for calculating it. Unfortunately, the largely 
fluctuating stochastic source as implied by the Noise Kernel for naked singular metric does not allow us to treat the backreaction as negligible. So, our 
results cannot easily be extended for studying the induced fluctuations on 
Cauchy Horizon. 

At the same time, we note that Noise Kernel is not very large till near the 
Cauchy Horizon (in the naked singularity case) and  so the background 
appears to be a good approximation despite backreaction.

Despite the above difficulty regarding the metric fluctuations, the source 
fluctuations can still be directly used for the purpose of interpreting.  
The semiclassical interpretation of the stress tensor divergence 
on Cauchy Horizon \cite{Sukratu} suggested that an energy burst on Cauchy
horizon is plausible. Such an interpretation would need to be reconsidered in 
view of our analysis. The highly fluctuating source near the Cauchy horizon 
would render it difficult to interpret any of its stochastic realizations as 
physically significant, in particular the semiclassical one of an energy burst 
on Cauchy Horizon.

The covered singularity case is in contrast with the above results. This is
similar to the difference seen in the semiclassical behavior of the
metric. In the latter analysis, the quantum stress tensor diverges
on the Cauchy horizon in the naked singular case against regular behavior
in the covered case. The contrast is borne out by analysis of fluctuations as 
well.

The divergence of quantum stress tensor in the semiclassical analysis has been
attributed to the fact that high curvature regions are exposed in the
naked singular case leading to high energy effects. It would seem that
high curvatures lead to diverging noise as well. Thus the contrast in the 
behavior of local quantum fields seems to be tied to the exposure of high 
curvature regions, or the lack of it. We suggest that this plausible connection could be investigated further.

Finally, the stochastic source fluctuating highly near the Cauchy Horizon,
could be interpreted for its effect as an {\it environment} (quantum fields) on the {\it system} (background metric) \cite{Hu1} . The spacetime could be highly sensitive to such {\it environmental} decoherence effects very near the Cauchy 
Horizon. In particular, as we have a quasi-free state, one could interpret 
this for example as spacetime reaction effects of particle creation. 
We suggest that this issue could be pursued separately.

\section*{Acknowledgements} 
We thank Inter University Centre for Astronomy and Astrophysics, Pune, India for providing
computing facilities and support. The authors are grateful to B. L. Hu, P. R. Anderson, A. Roura, J. Bates,
A. Eftekharzadeh and S.Sinha for useful discussions.


\begin{thebibliography}{99}
\bibitem{birrel}
 Birrell N. D. and Davies P. C. W.  \emph{Quantum fields in curved spacetime}
(Cambridge University Press, 1984). 
\bibitem{wald}
Robert Wald \emph{Quantum Field Theory in Curved Spacetime and Black Hole Thermodynamics}(University of Chicago Press, 1994).

\bibitem{visser}
Carlos Barcelo, Stefano Liberati, Sebastiano Sonego, Matt Visser, Phys.Rev.D {\textbf 77}:044032 (2008)

\bibitem{mattingly2}James Mattingly \textit{ Singularities and Scalar Fields.
Matter Theory and General Relativity.} Proceedings of the Seventh Biennial Meeting of the Philosophy of Science Association, 2000, Vancouver 

\bibitem{Sukratu}
Sukratu Barve, T.P.Singh Cenalo Vaz 

Phys.Rev D {\textbf 62}, 084021 (2000).\\ See also \\ Sukratu Barve, T.P.Singh, Cenalo Vaz and Louis Witten 
Phys.Rev. D {\textbf 58}, 104018 (1998). 



\bibitem{mattingly}James Mattingly \textit{Is quantum gravity necessary?} Proceedings of the 5th International Conference on the History and Foundations of General Relativity, 1999.


\bibitem{Hu1}
Bei Lok Hu and Enric Verdaguer, Living Rev. Relativity {\textbf 11}, 3 (2008).
\bibitem{Verd2}
E.Verdaguer 
XXIXth Spanish Meeting (ERE 2006), Journal of Physics: Conference Series {\textbf 66}, 012006 (2007).

\bibitem{Verd3}
R.Martin and E.Verdaguer , Phys. Rev D \textbf{60}, 084008 (1999).
\bibitem{Hu2}
N.G.Phillips and B.L.Hu, Phys Rev D \textbf{63}, 104001 (2001).


\bibitem{Sukanya}
Sukanya Sinha, Alpan Raval and B.L.Hu 
Foundations of Physics, Vol. 33, No. 1 January 2003.

\bibitem{FredenhagenHaag} Klaus Fredenhagen and Rudolf Haag
Comm. Math. Phys. \textbf{127}, No.2, 273 (1990). 

\bibitem{Nadal}
G.Perez-Nadal, A.Roura and E.Verdaguer, JCAP \textbf{05}, 036 (2010).
\bibitem{Fleming}
C.H.Fleming, A.Roura and B.L.Hu, 
arXiv:1004.1603.
\bibitem{Hu4}
N.G.Phillips and B.L.Hu, Phys. Rev D \textbf{67}, 104002 (2003). 
\bibitem{Roura}
A.Roura and E.Verdaguer, Int. J. Theor. Phys. \textbf{38}, 3123 (1999).
\bibitem{Roura2}
G.Perez-Nadal, A.Roura and E. Verdaguer, 
Journal of Cosmology and Astroparticle Physics. Volume 2010 (2009).
\bibitem{Cho}
H.T.Cho and B.L.Hu 
Journal of Physics: Conference series, Volume 330, conference 1.(2011).
\bibitem{Ifte}
A.Eftekharzadeh, Jason D.Bates, Albert Roura, Paul R. Anderson and B.L.Hu
Phys. Rev. D \textbf{85},044037 (2012).


\bibitem{Page}
D.N.Page, Phys. Rev D, \textbf{25}, 1499 (1982).

\bibitem{kineticqg} B. L. Hu, Int.J.Theor.Phys. {\textbf 41} 2091-2119 (2002)

\bibitem{kaywald}
Bernard S. Kay, Robert M. Wald, Phys. Rep. {\textbf 207}49-136 (1991)
\bibitem{radzikowski}
Marek J. Radzikowski, Comm. Math. Phys. {\textbf 179}529-553 (1996)



\bibitem{hu11}
N.G.Phillips and B.L.Hu 
Phys. Rev. D \textbf{63}, 104001 (2001).
\end{thebibliography}
\end{document}